# Multiferroic order parameters in rhombic antiferromagnets. RCrO$_3$

A.K. Zvezdin[1], Z.V. Gareeva[2], X.M. Chen[3]

[1] Prokhorov General Physics Institute, Russian Academy of Sciences, 119991, Moscow, Russia

[2] Institute of Molecule and Crystal Physics, Subdivision of the Ufa Federal Research Centre of the Russian Academy of Sciences, 450075, Ufa, Russia

[3] Laboratory of Dielectric Materials, School of Materials Science and Engineering, Zhejiang University, Zheda Road 38, 310027, Hangzhou, China

Corresponding authors: gzv@anrb.ru, zvezdin.ak@phystech.edu

## Abstract

Advanced multiferroic materials, useful in a variety of spintronic applications, memory technologies and neuromorphing computing, are of great interest. Currently, active research is aimed at a novel class of perovskite – based oxides, including rare earth orthoferrites and orthochromites, which exhibit magnetoelectric properties owed to intrinsic magnetic interactions in external electric and magnetic fields. Due to a variety of structural instabilities and couplings in these materials, understanding the underlying magnetoelectric mechanisms is still a challenge. In this paper, we explore magneteoelectricity of rare earth orthochromites from the symmetry point of view. We determine the principal structural order parameters and find their couplings with ferroelectric and magnetic orderings. Our calculations showed that electric dipole moments emerge in the vicinity of $Cr^{3+}$ ions in the unit cell of $RCrO_3$ due to the displacements of oxygen ions from their highly symmetric positions in the parent perovskite phase (structural instability). We find that the electric dipole moments are arranged in an antiferroelectric mode, so, in essence, $RCrO_3$ are antiferroelectric materials. By classifying the order parameters according to the irreducible representations of the $RCrO_3$ symmetry group ($D_{2h}^{16}$), we determine the possible couplings between distortive, ferroelectric and magnetic orderings and explore the emerging magnetoelectric structures in these terms. Our analysis makes it possible to explain experimentally observed polarization reversal and the concomitant reorientation of spins in a series of $RCrO_3$ compounds and to predict the possible scenarios of phase transitions in $RCrO_3$.

# 1. Introduction

Single-phase multiferroics are in the focus of modern physics due to the electric filed driven magnetism owed to cross-coupling effects that offer an efficient potential for fast and low - energy consuming spintronic technologies [1, 2]. Despite the diversity factors leading to multiferroic structures, including exchange – striction, *d-p* hybridization, lone – pairs coupling and other effects, current research on multiferroics is mainly focused on structure-controlled magnetoelectricity, implying that magnetism and ferroelectricity emerge from lattice strain effects and related crystallographic distortions.

In this respect, the materials with flexible crystal structure such as $ABO_3$ perovskite – based compounds are of the most interest. Intrinsic instability of the $ABO_3$ parent phase sensitive to the kind of cations occupying A and B positions offer wide opportunities for manipulating electronic structure and the properties related with magnetism, ferroelectricity, superconductivity etc. Most of perovskites crystallize in non - polar space group, so ferroelectricity is quite rare phenomena. The inclusion of magnetic ions Fe, Cr, Mn and R (rare earth ions) in perovskite structure leads to the emergence of antiferromagnetic (AFM) and weak ferromagnetic (WFM) properties as in the case of rare earth orthoferrites/orthochromites. As recent researches showed [3 - 9], AFM ordering in $RCr/FeO_3$ compounds induce improper ferroelectricity, which can be driven by electric field poling and external magnetic field [6].

Rare earth orthochromites belong to the family of rhombohedral antiferromagnets $RMO_3$, where *M* denotes the transition metal ions, *R = Y, La, Pr, Sm, Gd, Dy, Ho, Yb, Lu* stands for the rare earth, Lu or Y ions. Crystal structure and magnetic properties of rare earth orthochromites $RCrO_3$ studied since 1960s [10 - 12] are well established. They belong to the space symmetry group *Pbnm* ($D_{2h}^{16}$). Neutron diffraction measurements showed that the $Cr^{3+}$ ions order antiferromagnetically in G – type magnetic configurations with weak ferromagnetic component $\Gamma_1(A_x, G_y, C_z), \Gamma_2(F_x, G_z, C_y), \Gamma_4(F_z, A_y, G_x)$. Change of a temperature, doping, external effects induce the spin reorientation phase transitions (SRPT) between these states. The Neel temperature ($T_N^f$) of the rare earth sublattice being around of several *K* is much lower than $Cr^{3+}$ ordering temperature ($T_N^d$). As was shown in Refs. [11 - 13] $T_N^f$ strongly depends on the rare earth ion, its radii, configuration and ground state, e.g. $T_N^{Dy/Tb} \sim 3.8/5K$, $T_N^{Sm/Nd/Er} \sim 34/17/6K$. Spins of the rare earth ions can order and reorient due to the influence of the internal effective field of the $Cr^{3+}$ magnetic sublattice and external factors such as applied magnetic field and strains [13-15]. These local transitions give an impact in the spin reorientation processes and lead to exotic magnetization reversals [13, 16 - 19]

Though ferroelectricity in RCrO$_3$ is forbidden by symmetry ferroelectric behavior have been observed in the number of orthochromites above the antiferromagnetic ordering temperature $T_N$ [8]. As has been shown recently RCrO$_3$ electric polarization can achieve sufficiently high values of the order 0.5-0.7 μC/cm$^2$, however the physical origin of magnetoelectric effects in RCrO$_3$ remains under discussions. Emergence of electric polarization is explained in terms of structural transition from non – polar *Pbnm* into polar *Pmna* structural phase, the central – asymmetrical ordering of the *f* sublattices modes, inverse Dzyaloshinskii – Moriya and Heisenberg exchange interactions, disorder effects and coupling between electric dipole and magnetic moments of the rare earth ion [7, 8, 20 -22],

In this paper, we perform the symmetry consideration of magnetoelectric properties of RCrO$_3$ in view of structural instability and related crystallographic distortions. Using the data of neutron diffraction measurements, we determine the structural order parameters related with the oxygen octahedral rotations and the displacements of ferroelectric cations from the centrosymmetrical positions in perovskite parent phase. We calculate electric dipole moments and demonstrate their antiferroelectric arrangement in RCrO$_3$ unit cell. We perform the classification of the order parameters according to the irreducible representations (IRs) of the space symmetry group *Pnma* and determine interrelation between magnetic, ferroelectric and structural properties and the ways of their possible transformations.

## 2. Symmetry analysis

The unit cell of RMO$_3$ (Fig.1) contains 4 RMO$_3$ molecules, so it has 4 M$^{3+}$ and 4 R$^{3+}$ magnetic ions (in the case of R = Rare Earth Ion) located in the local positions differing by the symmetry of O$^{2-}$ environments. The *d* – ions (M$^{3+}$) occupy the positions 4*b*, the *f* – ions (R$^{3+}$) occupy the position 4*c*, oxygen ions occupy the positions 4*c* and 8*d* (in Wyckoff notation). Magnetic moments of the *d* – ions determined by the vectors **M**$_i$, (*i*=1-4) constitute 4 transition metal magnetic sublattices and the magnetic moments of the *f* – ions determined by vectors **m**$_i$ (*i*=1-4) constitute 4 rare earth magnetic sublattices. The combinations between magnetic moments of the *d* –and the *f* –sublattices determine magnetic modes $\boldsymbol{F} = \boldsymbol{M}_1 + \boldsymbol{M}_2 + \boldsymbol{M}_3 + \boldsymbol{M}_4$, $\boldsymbol{A} = \boldsymbol{M}_1 - \boldsymbol{M}_2 - \boldsymbol{M}_3 + \boldsymbol{M}_4$ $\boldsymbol{G} = \boldsymbol{M}_1 - \boldsymbol{M}_2 + \boldsymbol{M}_3 - \boldsymbol{M}_4, \boldsymbol{C} = \boldsymbol{M}_1 + \boldsymbol{M}_2 - \boldsymbol{M}_3 - \boldsymbol{M}_4$ of the '*d*' ions and magnetic modes $\boldsymbol{f} = \boldsymbol{m}_1 + \boldsymbol{m}_2 + \boldsymbol{m}_3 + \boldsymbol{m}_4$, $\boldsymbol{a} = \boldsymbol{m}_1 - \boldsymbol{m}_2 - \boldsymbol{m}_3 + \boldsymbol{m}_4$, $\boldsymbol{g} = \boldsymbol{m}_1 - \boldsymbol{m}_2 + \boldsymbol{m}_3 - \boldsymbol{m}_4$, $\boldsymbol{c} = \boldsymbol{m}_1 + \boldsymbol{m}_2 - \boldsymbol{m}_3 - \boldsymbol{m}_4$ of the '*f*'- ions. As neutron diffraction measurements showed RCrO$_3$ exhibit one of three *G* – type antiferromagnetic (AFM) configurations with weak magnetic component $\Gamma_1(A_x, G_y, C_z)$, $\Gamma_2$ ($F_x, C_y, G_z$), $\Gamma_4$ ($G_x A_y F_z$) whose preference depends on the temperature, type of the rare earth ion and external fields. Recent experiments [6] have demonstrated the possibility of electric field induced processes of spin reorientation, which indicates the presence of ferroelectric ordering in $RCrO_3$, but the underlying physical mechanisms responsible for these effects are not entirely clear.

In the present item, we aim to substantiate the origin of ferroelectricity from the point of view of symmetry analysis. At the first step, it is reasonable to assume that, as in the family of ferroelectric perovskites, ferroelectricity in $RCrO_3$ emerges because of dipole ordering caused by structural instabilities related with displacements of the $R^{3+}$ and $O^{2-}$ ions. To describe the transition from the parent perovskite phase into orthorhombic structure with *Pnma* space symmetry group, we introduce structural (distortive) order parameters determined by polar vectors $D_i$ and axial vectors $\Omega_i$. The polar vectors $D_i$ are attributed to electric dipole moments and the axial vectors $\Omega_i$. are related with rotation of oxygen octahedrons surrounding $Cr^{3+}$ ions. Below we consider these distortive order parameters in more details.

### 2.1. Polar distortions and order parameters $D_i$

To calculate the electric dipole moments in a frame of point charge model we determine the position of the electric dipole charge center as $r_q = \dfrac{\sum_i q_i r_{qi}}{\sum_i q_i}$ where $q_i$ are the signed magnitudes of the charges, $r_{qi}$ are the radius vectors of the charges in the local reference frame. For the perovskite – like compounds

$$r_q = \frac{\left(+\dfrac{3}{8}e\right)\cdot\sum_{i=1}^{8} r_R + \left(-\dfrac{2}{2}e\right)\cdot\sum_{i=1}^{6} r_O}{\left|8\cdot\left(+\dfrac{3}{8}e\right) + 6\cdot\left(-\dfrac{2}{2}e\right)\right|} \qquad r = (x, y, z) \tag{1}$$

where $e$ is the elementary charge, $r_R$ are the radius vectors of the rare earth ions, $r_O$ are the radius vectors of the oxygen ions measured from $Cr^{3+}$ ion. Neutronographic data showed that the position of $Cr^{3+}$ ion in orthochromites remains unchanged, so we choose $Cr^{3+}$ ion as the origin of the dipole moment and the origin of the local reference frame. Note that in the case of an ideal perovskite $ABO_3$ electric dipoles are absent since $r_q = 0$.

In the case of orthochromites, the electric dipole charge center deviates from the position of $Cr^{3+}$ ion due to the displacement of oxygen ions from their centrosymmetrical positions in the parent phase of perovskite (see Appendix A), hence $r_q \neq 0$. To find the coordinates of $R^{3+}$ and $O^{2-}$ ions surrounding each of $Cr^{3+}$ ion in a unit cell we consider the arrangement of symmetrically equivalent positions derived by applying of the symmetry operation of *Pnma* space group to the specific ion whose coordinates can be taken from neutronographic data or ab-initio calculations [24]. As follows from the calculations (Appendix A), the radius vectors of 4 electric dipoles $d_i = qr_i$ centered on $Cr^{3+}$ ions in the $RCrO_3$ unit cell differ from each other

$$r_1 = i\left(\frac{2y_1 + 2y_2 - 2x_2 + 1}{3} - \frac{1}{2}\right) + j\left(\frac{-2y_1 - 2y_2 - 2x_2 + 2}{3} - \frac{1}{2}\right) + k(-z_2)$$
$$r_2 = i\left(\frac{2y_1 + 2y_2 - 2x_2 + 1}{3} - \frac{1}{2}\right) + j\left(\frac{-2y_1 - 2y_2 - 2x_2 + 2}{3} - \frac{1}{2}\right) + k(z_2)$$
$$r_3 = i\left(\frac{-2y_1 - 2y_2 + 2x_2 - 1}{3} + \frac{1}{2}\right) + j\left(\frac{2y_1 + 2y_2 + 2x_2 + 1}{3} - \frac{1}{2}\right) + k(-z_2)$$
$$r_4 = i\left(\frac{-2y_1 - 2y_2 + 2x_2 - 1}{3} + \frac{1}{2}\right) + j\left(\frac{2y_1 + 2y_2 + 2x_2 + 1}{3} - \frac{1}{2}\right) + k(z_2)$$

(2)

where $\{x_1, y_1, z_1\}$ and $\{x_2, y_2, z_2\}$ are the coordinates of oxygen ions in 4*c* and 8*d* positions respectively.

So, each of the dipoles has its own orientation, so the ferroelectric ordering established in $RCrO_3$ is characterized by 4 ferroelectric sublattices with electric dipoles $d_i = 3er_i$. Emphasize that electric dipoles appear due to the outcome of oxygen ions from their high symmetry positions, the displacements of the $R^{3+}$ ions from their high symmetrical positions give no impact into the electric dipole moments in the first approximation (Appendix A). To find the basic ferroelectric vectors transforming on the irreducible representations (IR) of the *Pnma* space symmetry group, we consider the possible linear combinations between electric dipole moments

$$P = d_1 + d_2 + d_3 + d_4$$
$$Q_2 = d_1 - d_2 - d_3 + d_4$$
$$Q_3 = d_1 - d_2 + d_3 - d_4$$
$$D = d_1 + d_2 - d_3 - d_4$$

(3)

The arrangement of electric dipole moments in $RCrO_3$ obtained by use of eq. (2) is shown in Fig.1. It is seen that here $D$ vector attains the maximum value in contrast to $P$, $Q_{2,3}$ which are negligibly small. So, in $RCrO_3$ antiferroelectric structure ordered by $D$ mode is established (Appendix A, Fig.1).

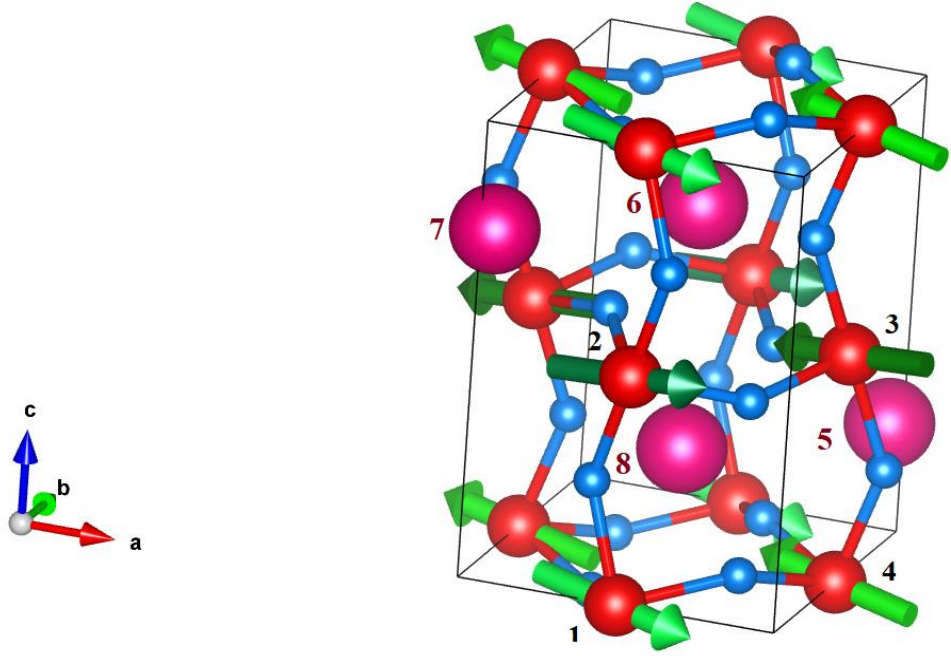

**Figure 1**. Electric dipole moments arrangement in RCrO$_3$ unit cell. Green arrows denote the orientation of electric dipole moments in the vicinity of Cr$^{3+}$ ions ordered by antiferroelectric ***D*** mode.

2.2. **Axial distortions and order parameters $\Omega_i$**.

Let us turn to crystallographic distortions associated with the rotation of the oxygen octahedra CrO6. To describe each of octahedrons surrounding Cr$^{3+}$ ions in a unit cell we introduce an axial vector $\boldsymbol{\omega}_i \parallel \boldsymbol{n}_{oi}$, where $\boldsymbol{n}_{oi}$ is the $i$ – the octahedron axis, $i$ is the number of Cr$^{3+}$ ion in the unit cell ($i$=1-4) (Fig.2).

Possible combinations between 4 vectors $\boldsymbol{\omega}_i$ are given in Table B (Appendix B). As known from experiments [10, 11], the strongest exchange interaction in RCrO$_3$ occurs between Cr-Cr ions arranged along $b \parallel [110]$ axis. This favors the orientation of the Dzyaloshinskii vector along this direction and implies the CrO6 octahedrons rotate around $b$ – axis. Accounting the G-type of AFM ordering in RCrO$_3$ we assume that the axial order parameter $\boldsymbol{\Omega}_b$ determined as

$$\boldsymbol{\Omega}_b = \boldsymbol{\omega}_1 - \boldsymbol{\omega}_2 + \boldsymbol{\omega}_3 - \boldsymbol{\omega}_4 \qquad (3)$$

attains the maximum values. To compare $|\boldsymbol{\Omega}_a| \ll |\boldsymbol{\Omega}_b|$, so we leave only $\boldsymbol{\Omega}_b$ for further consideration.

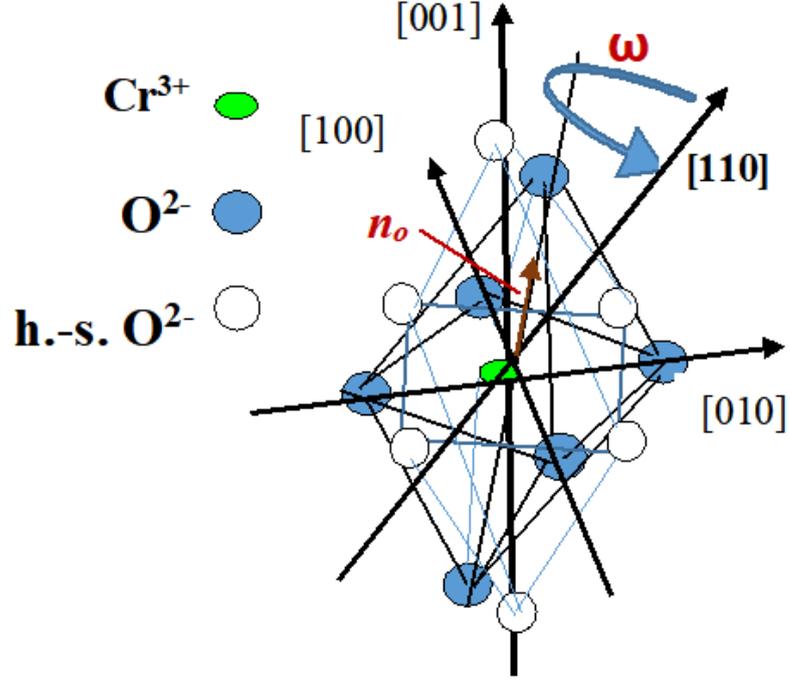

**Figure 2**. Schematic illustration of the rotation of the oxygen octahedron CrO6 around $b$- axis ($b \parallel [110]$), corresponding to the displacements of $O^{2-}$ ions (blue circles) from the highly symmetric (h.-s.) positions (white circles), here we introduce an axial vector $\omega$ directed along $n_o$, designated to describe the rotation of the ochahedron.

The arrangements of the assigned parameters ($d_i$, $\omega_i$) and the parity combinations of vectors $D$, $P$, $Q_i$, $\Omega_{a,b}$ transforming under the action of the generators of space group $Pnma$ $P\left(\bar{1}(+)2_x(+)2_y(+)2_z(+)\right), Q_2\left(\bar{1}(+)2_x(+)2_y(-)2_z(-)\right),...$ are listed in Table B (Appendix B).

## 2.3. Classification of the structural order parameters according to the irreducible representation of *Pnma* group

Using the parity combinations listed for every structural mode in Table B (Appendix B) we can classify the components of order parameters according to the irreducible representations (IRs) of *Pnma* symmetry group (Table 1). The *Pnma* symmetry group has 8 IRs, the first 4 of which ($\Gamma_1$-$\Gamma_4$) are time even and space – odd and the other ($\Gamma_5$-$\Gamma_8$) are time – odd and space - even. Magnetic modes related with $Cr^{3+}$ ions (**F, A, G, C**) and axial structural modes ($\Omega_a$, $\Omega_b$) belong to the time – even IR ($\Gamma_1$-$\Gamma_4$) while polarization (**P**), polar structural modes ($D_i$) and magnetic modes related with $R^{3+}$ ions (**f, a, g, c**) belong to the space even IRs ($\Gamma_5$-$\Gamma_8$).

*Table* 1. Irreducible representations of the *Pnma* symmetry group

| $\Gamma_i$ | $\bar{1}$ | $2_x$ | $2_y$ | $2_z$ | The components of basic magnetic order parameters and magnetic field | | The components of structural order parameters and electric field |
|---|---|---|---|---|---|---|---|
| | | | | | 4b | 4c | |
| $\Gamma_1$ | 1 | 1 | 1 | 1 | $A_x, G_y, C_z$ | $c_z$ | $\Omega_{by}$ |
| $\Gamma_2$ | 1 | 1 | -1 | -1 | $F_x, G_z, C_y, H_x$ | $f_x, c_y$ | $\Omega_{bz}$ |
| $\Gamma_3$ | 1 | -1 | 1 | -1 | $F_y, A_z, C_x, H_y$ | $f_y, c_x$ | - |
| $\Gamma_4$ | 1 | -1 | -1 | 1 | $F_z, A_y, G_x, H_z$ | $f_z$ | $\Omega_{bx}$ |
| $\Gamma_5$ | -1 | 1 | 1 | 1 | | $g_x, a_y$ | $Q_{2x}, Q_{3y}, D_z$ |
| $\Gamma_6$ | -1 | -1 | -1 | 1 | | $a_z$ | $P_x, Q_{3z}, D_y, E_x$ |
| $\Gamma_7$ | -1 | -1 | 1 | -1 | | $g_z$ | $P_y, Q_{2z}, D_x, E_y$ |
| $\Gamma_8$ | -1 | 1 | -1 | -1 | | $g_y, a_x$ | $P_z, Q_{2y}, Q_{3x}, E_z$ |

### 2.4. Magnetoelectric effect

Table 1 allows us to find the relations between ferroelectric and magnetic orderings and qualitatively analyze the possible couplings in the RCrO$_3$ crystal. Below we consider the manifestations of magnetoelectric effect, which is given by the dependences of polarization (***P***) on magnetic order parameters and magnetization (***F***) on the ferroelectric order parameters.

$$\begin{aligned}
\boldsymbol{P} = &\, \boldsymbol{i}\left(\kappa_x E_x + \eta_y D_y + \nu_z Q_{3z} + \beta_{xx} H_x g_x + \beta_{yy} H_y g_y + \beta_{zz} H_z g_z + \alpha_{zy} G_z a_y + \alpha_{zx} G_z g_x + \alpha_{xz} G_x g_z + \alpha_{yz} G_y a_z\right) + \\
&+ \boldsymbol{j}\left(\kappa_y E_y + \eta_x D_x + \nu_z Q_{2z} + \beta_{xy} H_x g_y + \beta_{yx} H_y g_x + \alpha_{zx} G_z a_x + \alpha_{zy} G_z g_y + \alpha_{xz} G_x a_z + \alpha_{yz} G_y g_z\right) + \\
&+ \boldsymbol{k}\left(\kappa_z E_z + \nu_y Q_{2y} + \nu_x Q_{3x} + \beta_{xz} H_x g_z + \beta_{zy} H_z g_y + \alpha_{zy} G_z a_y + \alpha_{zx} G_z g_x + \alpha_{xz} G_x g_z + \alpha_{yx} G_y a_x + \alpha_{yy} G_y g_y\right) + \ldots
\end{aligned}$$

(4)

***P*** vector in the first approximation are related with $Q_i$, ***D*** and electric field, in the second and higher approximations they can be expressed via magnetic moments of Cr$^{3+}$ and R$^{3+}$ ions. The higher order amendments are derived by use of multiplication rules for IRs (Appendix C).

The ferromagnetic vector ***F*** in terms of the magnetic and electric fields and distortive order parameters are written as follows

$$\begin{aligned}
\boldsymbol{F} = &\, \boldsymbol{i}\left(\chi_1 H_x + \alpha_{zy}\Omega_{bz} G_y + \gamma_{zz} g_z E_z + \gamma_{yy} g_y E_y + \gamma_{xy} a_x E_y + \gamma_{13} g_x E_x + \gamma_{yx} a_y E_x + b_{yx} D_y g_x + b_{xy} D_x g_y + \ldots\right) + \\
&+ \boldsymbol{j}\left(\chi_2 H_y + \alpha_{zxy}\Omega_{bz}\Omega_{bx} G_y + \gamma_{yx} g_y E_x + \gamma_{xx} a_x E_x + \gamma_{xy} g_x E_y + \gamma_{yy} a_y E_y + b_{yy} D_y g_y + b_{xx} D_x g_x + \ldots\right) + \\
&+ \boldsymbol{k}\left(\chi_3 H_z + \alpha_{xy}\Omega_{bx} G_y + \gamma_{zx} g_z E_x + \gamma_{zy} a_z E_y + \gamma_{xz} g_x E_z + \gamma_{yz} a_y E_z + b_{yz} D_y g_z + \ldots\right)
\end{aligned}$$

(5)

## 3. Poling effects, symmetry and transformations of magnetic and crystallographic structures due to electric field

In this section, we analyze the possible effects caused by electric field applied to a sample. Electric poling can induce polar ordering accompanied with emergence of electric polarization and transformation of magnetic and crystallographic structures. The possible structures and phase transitions can be understood from the symmetry consideration.

Let's appeal to Table 1 of the IRs of *Pnma* symmetry group. All order parameters belonging to the same IR compose the definite magnetic, ferroelectric and crystallographic structures, i.e. every atomic, dipolar and magnetic configuration corresponds to its own IR. That means that the transition from one IR to another initiated by external factor, in our case polishing the sample with electric field, manifests itself in structural and magnetic transformations determined by specific symmetry operations. Note that reorientation phase transitions, in which exchange-coupled structures are preserved, can also occur within the same IR. Using Table 1, we extract the symmetry elements, magnetic and point symmetry group corresponding to each of the IR of *Pnma* group.

Consider IR $\Gamma_1$. Here, all symmetry elements $\bar{1}, 2_x, 2_y, 2_z$ remain the magnetic symmetry elements and compose magnetic point symmetry group $\bar{1}2_x2_y2_z = m_x m_y m_z = mmm$ written in terms of symmetry planes $m = \bar{1}\cdot 2$. Turn to IR $\Gamma_2$. After magnetic ordering here, the inversion operation $\bar{1}$ and the axis $2_x$ remain symmetry elements while the axes $2_y$, $2_z$ change the sign of magnetic components, so they are replaced by $\bar{2}_y, \bar{2}_z$ elements. The renewed set of symmetry elements determine $m_x\bar{m}_y\bar{m}_z = m\bar{m}\bar{m}$ magnetic point symmetry group and point group $C_{2h}$. To obtain all symmetry elements of a point group we use the relations defining combinations of symmetry operations such as

$$\bar{2}_y \oplus \bar{2}_z = \bar{2}_x = \bar{1} \oplus 2_x = \sigma_x, \bar{1} \oplus 2_y \oplus 2_z = \bar{1} \oplus 2_x = \sigma_x, T \oplus \bar{2}_y \oplus \bar{2}_z = T \oplus \bar{2}_x = T \oplus \sigma_x = R\sigma_x.$$

Consider magnetic structures and atomic arrangements whose OPs transform by the same IR.

Take as an example the IR $\Gamma_2$. $G_z$–type AFM order or weak FM (WFM) order $(F_x, G_z, C_y)$ are invariant to the symmetry operations of the corresponding magnetic group $m\bar{m}\bar{m}$. Also, the $z$ – component of the axial vector $\Omega_b$ that determines the arrangement of the CO6 octahedrons is transformed by the IR $\Gamma_2$. The same situation is realized for the IRs $\Gamma_4$ and $\Gamma_1$. So, in the case of $\Gamma_4$ magnetic structures form the $G_x$–type AFM order and weak FM (WFM) order $(F_z, G_x, A_y)$ which transform according to IR $\Gamma_4$ as well as the $x$– component of the axial vector $\Omega_b$. In the case of $\Gamma_1$ magnetic structures form $G_y$ –type AFM order and weak FM (WFM) order $(A_x, G_y, C_z)$ which as well as the $y$– component of axial vector $\Omega_b$ transform according to the IR $\Gamma_1$. Thus, one can conclude that the $G$–type AFM order and WFM states are stabilized due to the CO6 octahedron rotation, since the components of axial parameter $\Omega_b$ as well as the corresponding components of the principle magnetic parameter $G$ transform according to the same IRs.

By its nature, the electric field ($E$) can not directly affect the magnetization, but $E$ can influence the atomic arrangements and the related structural order parameters. Consider the transitions induced by an electric field from the point of view of symmetry. We start with transitions within one IR that do not change the electric dipole coupled structures (ECS). So, the antiferroelectric dipole arrangement determined by the $y$ – component of $D$ vector (Fig. 3a) can be reoriented into ferroelectric ECS characterized by the $x$– component of $P$ vector (Fig.3b) in the electric field $E$ applied along the $a$ – axis since they belong to the IR $\Gamma_6$. In its turn, these ECS are coupled with magnetic configurations that transform according to $\Gamma_6$ (Table 1) which are $g_x G_z$; $a_y G_z$; $g_z G_x$. Thus, we expect that the ferroelectric phase transition (FPT) $D_y \rightarrow P_x$ is accompanied with spin reorientation phase transition (SRPT) $g_z G_x \rightarrow g_x G_z$ (Fig.3).

Considering IR $\Gamma_7$ one can conclude that antiferroelectric ECS with $D_x$ is reoriented into ferroelectric ECS with $P_y$ in the electric field $E$ applied along $b$ – axis, the corresponding magnetic SRPT is $g_y G_z \rightarrow a_z G_x$ (here we take into account the orthogonality between magnetic and ferroelectric vectors).

Transitions accompanied by the destruction of the ECS, that is, transitions between ferroelectric structures transforming through different IRs, is expected to occur in a stronger electric field. For example, consider the transition $D_y \rightarrow P_y$ ($\Gamma_6 \rightarrow \Gamma_7$). This type of FPT should be accompanied with a spin reorientation from the magnetic state determined by one of the combinations $g_x G_z$; $a_y G_z$; $g_z G_x$ ($\Gamma_6$) into magnetic state determined by one of the combinations $g_y G_z$; $a_z G_x$ ($\Gamma_7$) such as SRPT $g_x G_z \rightarrow g_y G_z$; $g_z G_x \rightarrow g_y G_z$. In the case of FPT $D_x \rightarrow P_x$ ($\Gamma_7 \rightarrow \Gamma_6$) the SRPT $g_y G_z \rightarrow g_x G_z$; $g_y G_z \rightarrow g_z G_x$ occur, here we also use the assumption of orthogonality of dipole and magnetic moments.

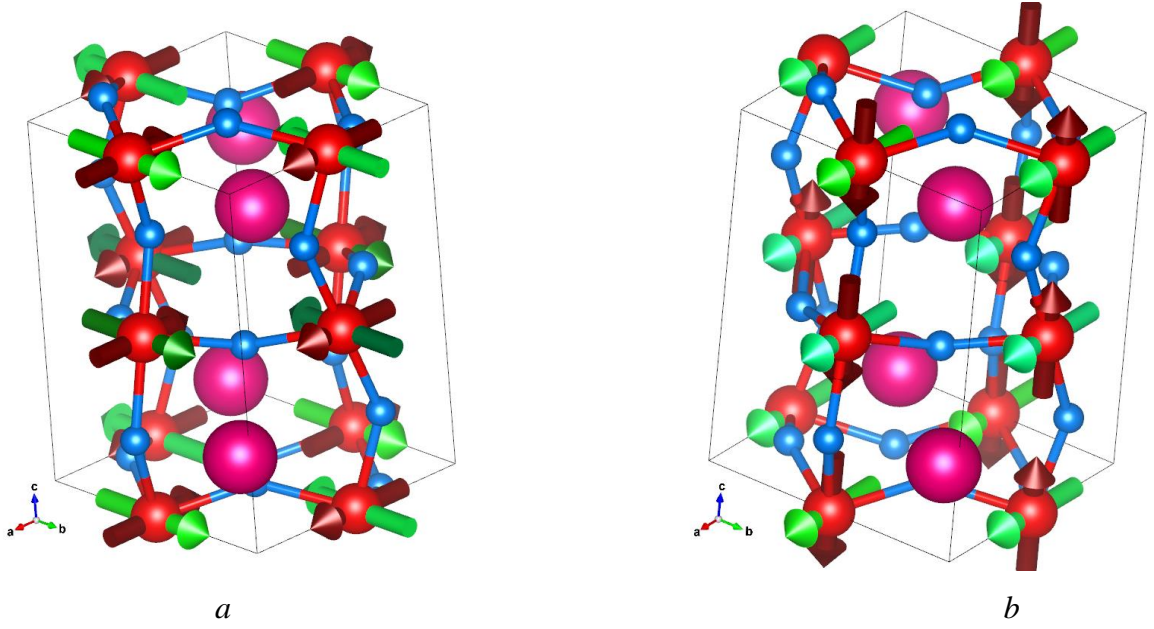

*a*  *b*

**Figure 3**. Electric dipole coupled structures (ECS) transforming according to the IR $\Gamma_6$, ferroelectric phase transition $D_y \rightarrow P_x$ in electric field **E** applied along the *a* – axis. Corresponding spin reorientation phase transition is $g_z G_x \rightarrow g_x G_z$, a) arrangements of electric dipole moments (green arrows) in $D_y$ – mode, magnetic moments of $Cr^{3+}$ ions (red arrows) in $G_x$– mode, b) arrangement of electric dipole moments in $P_x$ – mode, magnetic moments of $Cr^{3+}$ ions (red arrows) in $G_y$– mode

So electric polishing along the **OX**||*a*- axis induces AFM $G_z$ or WFM $\left(F_x, G_z, C_y\right)$ order, which transforms according to the IR $\Gamma_2 = \Gamma_6 \otimes \Gamma_5$. Polishing in an electric field applied along the **OY**||*b*- axis ($E_y \in \Gamma_7$) stabilize the magnetic configuration $G_x$, which transforms according to the IR $\Gamma_4 = \Gamma_8 \otimes \Gamma_5$ (here we consider the FPT inside one IR).

## 4. Discussion

In the frame of symmetry consideration, let us analyze $RCrO_3$ magnetoelectric properties exhibited during spin reorientation phase transitions. As an example we consider the transition between $\Gamma_4$ ($G_x$, $A_y$, $F_z$), $\Gamma_2$($F_x$, $C_y$, $G_z$), $\Gamma_1\left(A_x, G_y, C_z\right)$ magnetic states induced by the temperature. These transitions have been explored in $RCrO_3$ in Ref. [6] for several compositions $RCrO_3$ (R=Sm, Tm, Tb, Gd, Er, Lu).

1. Consider $TmCrO_3$, the measured spin - flop transition and the associated drop of polarization are shown in Fig.4.

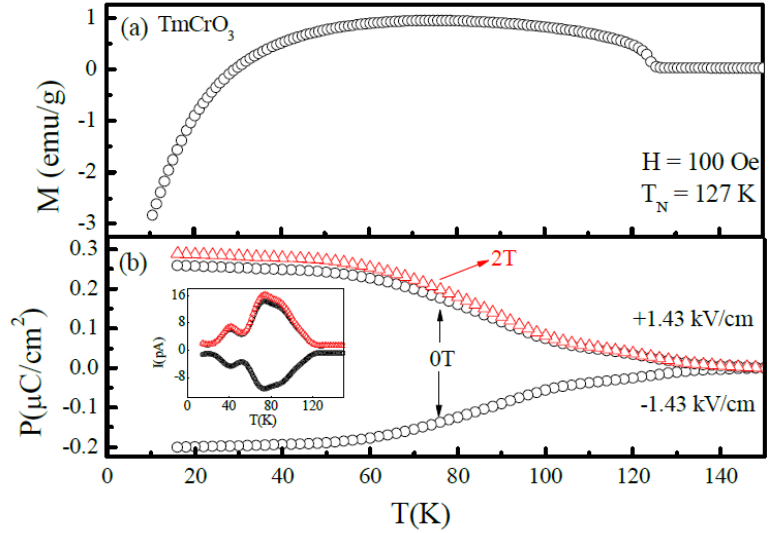

**Figure 4**. a) Field-cooled magnetization of TmCrO$_3$ measured at 100 Oe showing temperature induced magnetization reversal. b) Electric polarization (corrected for leakage) as a function of temperature. Poling is done at two different fields +1.43 kV/cm and -1.43 kV/cm (black) at 0 and 2T field (red). Inset in (b) shows the pyroelectric current as a function of temperature at two different poling fields with (red) and without the presence of magnetic field of 2T (black) [6].

As can be seen in Fig.4, the change in polarization that occurs at $T\sim80\,K$ is not accompanied by a change in the magnetization (spin reorientation). From another side, at the $T\sim30$ K the change of magnetization occurs and polarization does not change. Magnetic properties of TmCrO3 have been studied experimentally in Ref. [31-33]. In the absence of magnetic field at $T<5.6$ K from the high – temperature phase $\Gamma_2$ into the low temperature phase $\Gamma_4$ [31], the magnetization reversal at $T\sim28$ K has been reported quite recently [32, 33]. So, we can conclude that the magnetization reversal shown in Fig.4 at $T\sim28$ K is due to magnetic couplings and analyze the ferroelectric transition at $T\sim80$ K. In weak ferromagnets magnetization is determined by ferromagnetic vector **F**. So we should consider ferroelectric phase transition (FPT) within the IR $\Gamma_2$ phase with $F_z$ – component supported by field **H** (Fig. 4). According to multiplication rules (Appendix C) the IR $\Gamma_2$ can be represented as

$$\Gamma_2 = \Gamma_8 \otimes \Gamma_7 = \Gamma_2(P_z, g_z) \qquad (7)$$

Thus, in terms of the IRs of *Pnma* group (Table 1) the transition in TmCrO$_3$ shown in Fig.4 is described as follows

$$\Gamma_2(F_x, C_y, G_z) \to \Gamma_2(F_x, C_y, G_z)$$
$$\Gamma_2(P_z g_z) \to \Gamma_2(D_x g_y = 0) \qquad (8)$$

2. Consider GdCrO$_3$, the magnetic and ferroelectric transitions during cooling (Ref. [6]) occurring at the temperature $T$~150 $K$ ($M$~0 $emu/g$) are shown in Fig.5. Since even in **H**||**OZ** |**F**|~0 we have the magnetic state $\Gamma_2(F_x, G_z, C_y)$ and FPT within the IR $\Gamma_2$. So, the transition in GdCrO$_3$ shown in Fig.5 is described as follows

$$\Gamma_2(F_x, G_z, C_y) = \Gamma_2(F_x, G_z, C_y)$$
$$\Gamma_2(P_x g_x) \to \Gamma_2(D_z a_z = 0) \tag{9}$$

since

$$\Gamma_2 = \Gamma_6 \otimes \Gamma_5 = \Gamma_2(P_x, a_y / g_x), \quad \Gamma_2 = \Gamma_7 \otimes \Gamma_8 = \Gamma_2(P_z, g_z) = \Gamma_2(P_y, g_y / a_x) \tag{10}$$

Note that FPT $\Gamma_2(P_y g_y) \to \Gamma_2(D_z a_z = 0)$ can also occur.

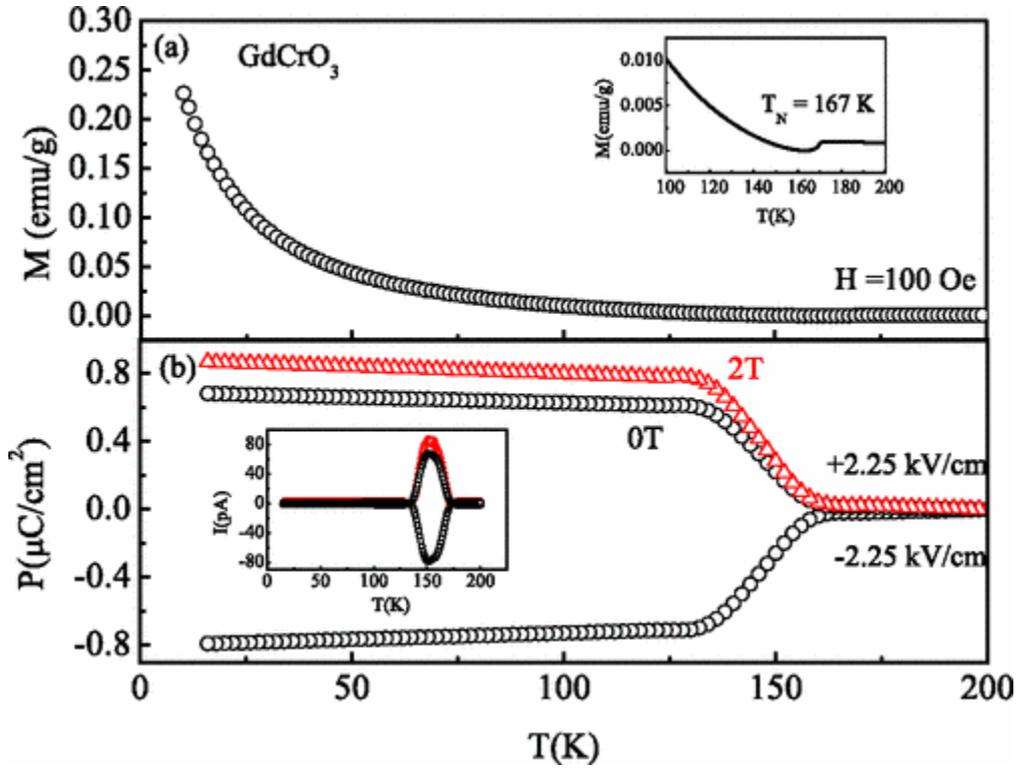

**Figure 5**. Effect of applied magnetic field on electric polarization in GdCrO$_3$ [6]. It can be seen that the polarization changes systematically with the strength and polarity of the applied magnetic field.

3. Consider EuCrO$_3$, the SRPT during cooling (Ref. [6]) at $T$~20 $K$ shown in Fig.6 is described as

$$\Gamma_1(A_x, G_y, C_z) \to \Gamma_4(F_z, A_y, G_x) \tag{11}$$

The corresponding FPT can be described as

$$\Gamma_1(D_z D_z) \to \Gamma_4(P_x g_z) \tag{12}$$

Since $\Gamma_1 = (\Gamma_i)^2$

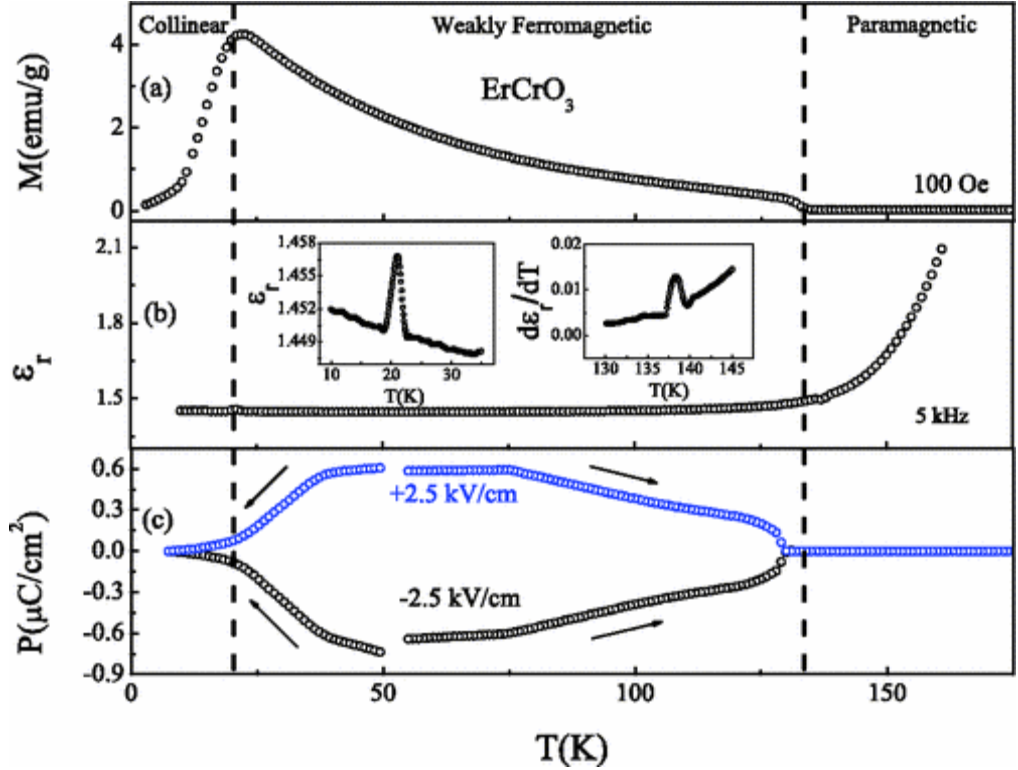

**Figure 6.** a) Field cooled magnetization of ErCrO$_3$ at 100 Oe showing a spin reorientation transition from weakly ferromagnetic $\Gamma_4$ to collinear $\Gamma_1$ magnetic spin structure at $T\sim22\ K$, b) dielectric constant as a function of temperature at 5 kHz. Insets show the dielectric anomaly at both the TSR and $T_N$ temperature regions, c) ferroelectric polarization as a function of temperature [6].

The similar situation develops in SmCrO$_3$ which exhibits ferroelectric features at the magnetic ordering temperature of iron ($T_{NFe} = 670\ K$). The drop in magnetization below 40 $K$ due to spin reorientation is accompanied with the reorientation of polarization.

To describe this transition, we choose the suitable combinations of the corresponding IRs obtained by use of Table 1 and the IRs multiplications shown in Table C.

To summarize, we demonstrate that rare earth orthochromites have intrinsic multiferroic properties.

## 5. Conclusion

To conclude, we analyze the correlation between structural instabilities and magnetoelectric properties of RCrO$_3$ in the framework of symmetry analysis. We showed that electric dipoles emerge in the RCrO$_3$ compounds, whose crystal structure belongs to the *Pnma* space symmetry group, due to the rotation of the CrO$_6$ octahedrons and the related displacements of oxygen ions from their centro- symmetrical positions in the perovskite praphase. In an absence of external agents, such as electric field, the electric dipoles in the rare earth orthochromites are arranged in an antiferroelectric *D* –mode. The polishing of the RCrO$_3$ crystals with external electric field will lead to reordering of the electric dipoles and emerging of an electric polarization, which was experimentally observed in Refs. [6, 29 - 33]. Using methods of theory group analysis, we classify structural order parameters (the axial distortive vector **Ω$_b$** and the polar vectors *D, Q$_{2,3}$, P*) along with magnetic ones (*F*, *A*, *G*, *C*) according to the irreducible representations of the RCrO$_3$ symmetry group; determine the magneto-electric-distortive couplings; magnetoelectric effect the possible scenarios of phase transitions. We analyze transformations of magnetic and ferroelectric states and compare our findings with the results of experimental measurements of the spin reorientation processes accompanied with polarization reversal in the series of RCrO$_3$ (R=Sm, Tm, Tb, Gd, Er, Lu) crystals polished with electric field [6].

## Acknowledgements

The work was supported by the Russian Foundation for Basic Research (grant No. 19-52-80024).

**Appendix A. Coordinates of electric dipoles in RCrO3 unit cell**

As an example, we considered YCrO$_3$ [24] whose Wyckoff coordinates are given in Table 1, for convenience we bind the coordinates with specific ions (Fig.1) whose numbers are indicated in the Table A1.

Table A1. Wyckoff coordinates for *Pnma* space group

| YCrO3 Ion{multiplicity Wyckoff letter} | Coordinate values (T=321 K) [24] | Coordinates of the ions |
|---|---|---|
| Y{4c} | (0.0169, 0.25, 0.0689) $\left(x_R, \frac{1}{4}, y_R\right)$ | $\left(x_R, \frac{1}{4}, y_R\right), \left(-x_R, \frac{3}{4}, -y_R\right),$ $\left(-x_R+\frac{1}{2}, \frac{3}{4}, y_R+\frac{1}{2}\right), \left(x_R+\frac{1}{2}, \frac{1}{4}, -y_R+\frac{1}{2}\right)$ |
| O1{4c} | (0.105, 0.25, 0.46) $\left(x_1, \frac{1}{4}, y_1\right)$ | $\left(x_1, \frac{1}{4}, y_1\right), \left(-x_1, \frac{3}{4}, -y_1\right), \left(-x_1+\frac{1}{2}, \frac{3}{4}, y_1+\frac{1}{2}\right),$ $\left(x_R+\frac{1}{2}, \frac{1}{4}, -y_R+\frac{1}{2}\right)$ |
| Cr{4b} | $\left(\frac{1}{2}, 0, 0\right)$ | $\left(0, 0, \frac{1}{2}\right), \left(0, \frac{1}{2}, \frac{1}{2}\right), \left(\frac{1}{2}, \frac{1}{2}, 0\right), \left(\frac{1}{2}, 0, 0\right)$ |
| O2{8d} | (-0.3082, 0.0545, 0.3039) $(x_2, z_2, y_2)$ | $(x_2, z_2, y_2), \left(-x_2+\frac{1}{2}, -z_2, y_2+\frac{1}{2}\right), \left(-x_2, z_2+\frac{1}{2}, -y_2\right),$ $\left(x_2+\frac{1}{2}, -z_2+\frac{1}{2}, -y_2+\frac{1}{2}\right), \left(x_2, -z_2+\frac{1}{2}, y_2\right),$ $\left(-x_2+\frac{1}{2}, y_2+\frac{1}{2}, z_2+\frac{1}{2}\right),$ $(-x_2, -z_2, -y_2), \left(x_2+\frac{1}{2}, z_2, -y_2+\frac{1}{2}\right)$ |

where $\{x_R, y_R, z_R\}$ and $\{x_1, y_1, z_1\}$, $\{x_2, y_2, z_2\}$ are the coordinates of Y$^{3+}$ (R$^{3+}$) ions in 4c positions, and O$^{2-}$ ions in 4c, 8d positions respectively.

In the following we change $z \to y$ and transform to the pseudocubic coordinate frame (OX'||[100]. OY'||[010]. OZ'||[001]) $x' = x/\sqrt{2} - y/\sqrt{2}$, $y' = x/\sqrt{2} + y/\sqrt{2}$. Below we write the

coordinate of the nearest neighboring ions surrounding each of $Cr^{3+}$ ion in the unit cell of RCrO3 which were obtained by use of the symmetry operation of *Pnma* space group to the specific ion (Table A1) along with translations. So, we consider 4 cubes with $Cr^{3+}$ ions in their centers.

1) Consider $Cr^{3+}$ (1) with coordinates $\left(\frac{1}{2},\frac{1}{2},0\right)$ in pseudocubic reference frame. The coordinates of 8 $Y^{3+}$ ($R^{3+}$) and 6 $O^{2-}$ ions surrounding $Cr^{3+}$ (1) are given in Table 2A, 3A.

Table 2A. The coordinates of $Y^{3+}$ ($R^{3+}$) ions surrounding $Cr^{3+}$ (1).

| Y/R | 8 | 5 | 8-$1_{y'}$ | 5-$1_{y'}$ | 6-$1_{z'}$ | 7-$1_{z'}$ | 6-$1_{y'}$-$1_{z'}$ | 7-$1_{y'}$-$1_{z'}$ |
|---|---|---|---|---|---|---|---|---|
| x | $x_R+\bar{y}_R$ | $x_R+y_R$ | $x_R+\bar{y}_R$ | $x_R+y_R$ | $\bar{x}_R+\bar{y}_R$ | $\bar{x}_R+y_R$ | $\bar{x}_R+\bar{y}_R$ | $\bar{x}_R+y_R$ |
| y | $x_R+y_R$ | $x_R+\bar{y}_R+1$ | $x_R+y_R-1$ | $x_R+\bar{y}_R$ | $\bar{x}_R+y_R+1$ | $\bar{x}_R+\bar{y}_R$ | $\bar{x}_R+y_R$ | $\bar{x}_R+\bar{y}_R-1$ |
| z | 1/4 | 1/4 | 1/4 | 1/4 | -1/4 | -1/4 | -1/4 | -1/4 |

Table 3A. The coordinates of $O^{2-}$ ions surrounding $Cr^{3+}$ (1).

| $O^{2-}$ | $4_{O1}$ | $3_{O1}$-$1_{z'}$+$1_{x'}$ | $5_{O2}$ | $6_{O2}$ | $5_{O2}$+$1_{x'}$ | $2_{O2}$-$1_{y'}$-$1_{x'}$ |
|---|---|---|---|---|---|---|
| x | $x_1+y_1$ | $\bar{x}_1+y_1+1$ | $\bar{x}_2+y_2$ | $x_2+y_2$ | $\bar{x}_2+y_2+1$ | $\bar{x}_2+\bar{y}_2-1$ |
| y | $x_1+\bar{y}_1+1$ | $\bar{x}_1+\bar{y}_1$ | $\bar{x}_2+\bar{y}_2$ | $x_2+\bar{y}_2+1$ | $\bar{x}_2+\bar{y}_2$ | $\bar{x}_2+y_2$ |
| z | 1/4 | -1/4 | $-z_2$ | $z_2$ | $-z_2$ | $-z_2$ |

Calculate the coordinates of the charge center in the 1st cube implementing eq. (2)

$$x_{Cr1}=\frac{2y_1+2y_2-2x_2+1}{3}, y_{Cr1}=\frac{-2y_1-2y_2-2x_2+2}{3}, z_{Cr1}=-z_2$$

2) Consider $Cr^{3+}$ (2) with coordinates $\left(\frac{1}{2},\frac{1}{2},\frac{1}{2}\right)$. The coordinates of 8 $Y^{3+}$ ($R^{3+}$) and 6 $O^{2-}$ ions surrounding $Cr^{3+}$ (2) are given in Table 4A, 5A.

Table 4A. The coordinates of $Y^{3+}$ ($R^{3+}$) ions surrounding $Cr^{3+}$ (2).

| Y/R | 8 | 5 | 6 | 7 | 5-$1_{y'}$ | 8-$1_{y'}$ | 6-$1_{y'}$ | 7-$1_{y'}$ |
|---|---|---|---|---|---|---|---|---|
| x | $x_R+\bar{y}_R$ | $x_R+y_R$ | $\bar{x}_R+\bar{y}_R$ | $\bar{x}_R+y_R$ | $x_R+y_R$ | $x_R+\bar{y}_R$ | $\bar{x}_R+\bar{y}_R$ | $\bar{x}_R+y_R$ |
| y | $\bar{x}_R+\bar{y}_R$ | $x_R+\bar{y}_R+1$ | $\bar{x}_R+y_R+1$ | $\bar{x}_R+\bar{y}_R$ | $x_R+\bar{y}_R$ | $\bar{x}_R+\bar{y}_R-1$ | $\bar{x}_R+y_R$ | $\bar{x}_R+\bar{y}_R-1$ |
| z | 1/4 | 1/4 | 3/4 | 3/4 | 1/4 | 1/4 | 3/4 | 3/4 |

Table 5A. The coordinates of $O^{2-}$ ions surrounding $Cr^{3+}$ (2).

| $O^{2-}$ | $4_{O1}$ | $3_{O1}+1_{x'}$ | $3_{O2}$ | $6_{O2}$ | $3_{O2}+1_{x'}$ | $2_{O2}-1_{y'}-1_{x'}$ |
|---|---|---|---|---|---|---|
| x | $x_1 + y_1$ | $\bar{x}_1 + y_1 + 1$ | $\bar{x}_2 + y_2$ | $x_2 + y_2$ | $\bar{x}_2 + y_2 + 1$ | $\bar{x}_2 + \bar{y}_2 - 1$ |
| y | $x_1 + \bar{y}_1 + 1$ | $\bar{x}_1 + \bar{y}_1$ | $\bar{x}_2 + \bar{y}_2$ | $x_2 + \bar{y}_2 + 1$ | $\bar{x}_2 + \bar{y}_2$ | $\bar{x}_2 + y_2$ |
| z | 1/4 | 3/4 | $z_2 + \dfrac{1}{2}$ | $z_2$ | $z_2 + \dfrac{1}{2}$ | $-z_2$ |

Calculate the coordinates of the charge center in the 2$^{nd}$ cube

$$x_{Cr2} = \frac{2y_1 + 2y_2 - 2x_2 + 1}{3}, y_{Cr2} = \frac{-2y_1 - 2y_2 - 2x_2 + 2}{3}, z_{Cr2} = z_2 + \frac{1}{2}$$

3) Consider $Cr^{3+}$ (3) $\left(-\dfrac{1}{2}, \dfrac{1}{2}, \dfrac{1}{2}\right)$. The coordinates of 8 $Y^{3+}$ ($R^{3+}$) and 6 $O^{2-}$ ions surrounding $Cr^{3+}$ (3) are given in Table 6A, 7A.

Table 6A. The coordinates of $Y^{3+}$ ($R^{3+}$) ions surrounding $Cr^{3+}$ (3).

| Y/R | 8 | 5 | 6 | 7 | 5+1$_{y'}$ | 8+1$_{y'}$ | 6+1$_{y'}$ | 7+1$_{y'}$ |
|---|---|---|---|---|---|---|---|---|
| x | $x_R + \bar{y}_R$ | $x_R + y_R$ | $\bar{x}_R + \bar{y}_R$ | $\bar{x}_R + y_R$ | $x_R + y_R$ | $x_R + \bar{y}_R$ | $\bar{x}_R + \bar{y}_R$ | $\bar{x}_R + y_R$ |
| y | $\bar{x}_R + \bar{y}_R$ | $x_R + \bar{y}_R + 1$ | $\bar{x}_R + y_R + 1$ | $\bar{x}_R + \bar{y}_R$ | $x_R + \bar{y}_R + 2$ | $\bar{x}_R + \bar{y}_R + 1$ | $\bar{x}_R + y_R + 2$ | $\bar{x}_R + \bar{y}_R + 1$ |
| z | 1/4 | 1/4 | 3/4 | 3/4 | 1/4 | 1/4 | 3/4 | 3/4 |

Table 7A. The coordinates of $O^{2-}$ ions surrounding $Cr^{3+}$ (3).

| $O^{2-}$ | $1_{O1}$ | $2_{O1}-1_{x'}$ | $7_{O2}$ | $4_{O2}$ | $8_{O2}+1_{y'}-1_{x'}$ | $7_{O2}+1_{x'}$ |
|---|---|---|---|---|---|---|
| x | $x_1 + \bar{y}_1$ | $\bar{x}_1 + \bar{y}_1 - 1$ | $x_2 + \bar{y}_2$ | $x_2 + y_2$ | $\bar{x}_2 + \bar{y}_2 - 1$ | $x_2 + \bar{y}_2 + 1$ |
| y | $x_1 + y_1$ | $\bar{x}_1 + y_1 + 1$ | $x_2 + y_2$ | $x_2 + \bar{y}_2 + 1$ | $\bar{x}_2 + y_2 + 2$ | $x_2 + y_2$ |
| z | 1/4 | 3/4 | $-z_2 + \dfrac{1}{2}$ | $-z_2 + \dfrac{1}{2}$ | $z_2 + \dfrac{1}{2}$ | $-z_2 + \dfrac{1}{2}$ |

Calculate the coordinates of the charge center in the 3$^{rd}$ cube

$$x_{Cr3} = \frac{-2y_1 - 2y_2 + 2x_2 - 1}{3}, y_{Cr3} = \frac{2y_1 + 2y_2 + 2x_2 + 1}{3}, z_{Cr3} = -z_2 + \frac{1}{2}$$

4) Consider $Cr^{3+}$ (4) $\left(-\frac{1}{2}, \frac{1}{2}, 0\right)$. The coordinates of 8 $Y^{3+}$ ($R^{3+}$) and 6 $O^{2-}$ ions surrounding $Cr^{3+}$ (4) are given in Table 8A, 9A.

Table 8A. The coordinates of $Y^{3+}$ ($R^{3+}$) ions surrounding $Cr^{3+}$ (4).

| Y/R | 8 | 5 | 8+1$_{y'}$ | 5+1$_{y'}$ | 6-1$_{z'}$ | 7-1$_{z'}$ | 6+1$_{y'}$-1$_{z'}$ | 7+1$_{y'}$-1$_{z'}$ |
|---|---|---|---|---|---|---|---|---|
| x | $x_R + \bar{y}_R$ | $x_R + y_R$ | $x_R + \bar{y}_R$ | $x_R + y_R$ | $\bar{x}_R + \bar{y}_R$ | $\bar{x}_R + y_R$ | $\bar{x}_R + \bar{y}_R$ | $\bar{x}_R + y_R$ |
| y | $x_R + y_R$ | $x_R + \bar{y}_R + 1$ | $x_R + y_R + 1$ | $x_R + \bar{y}_R + 2$ | $\bar{x}_R + y_R + 1$ | $\bar{x}_R + \bar{y}_R$ | $\bar{x}_R + y_R + 2$ | $\bar{x}_R + \bar{y}_R + 1$ |
| z | 1/4 | 1/4 | 1/4 | 1/4 | -1/4 | -1/4 | -1/4 | -1/4 |

Table 9A. The coordinates of $O^{2-}$ ions surrounding $Cr^{3+}$ (4).

| $O^{2-}$ | 1$_{O1}$ | 2$_{O1}$-1$_{x'}$-1$_{z'}$ | 1$_{O2}$ | 1$_{O2}$+1$_{x'}$ | 6$_{O2}$ | 2$_{O2}$-1$_{x'}$+1$_{y'}$ |
|---|---|---|---|---|---|---|
| x | $x_1 + \bar{y}_1$ | $\bar{x}_1 + \bar{y}_1 - 1$ | $x_2 + \bar{y}_2$ | $x_2 + \bar{y}_2 + 1$ | $x_2 + y_2$ | $\bar{x}_2 + \bar{y}_2 - 1$ |
| y | $x_1 + y_1$ | $\bar{x}_1 + y_1 + 1$ | $x_2 + y_2$ | $x_2 + y_2$ | $x_2 + \bar{y}_2 + 1$ | $\bar{x}_2 + y_2 + 2$ |
| z | 1/4 | -1/4 | $z_2$ | $z_2$ | $z_2$ | $-z_2$ |

Calculate the coordinates of the charge center in the 4$^{th}$ cube

$$x_{Cr4} = \frac{-2y_1 - 2y_2 + 2x_2 - 1}{3}, \quad y_{Cr4} = \frac{2y_1 + 2y_2 + 2x_2 + 1}{3}, \quad z_{Cr4} = z_2$$

**Appendix B. Axial and polar order parameters of RCrO3 and their transformation on the symmetry operation of *Pnma* space group**

Here we show possible arrangements of the structural parameters ($d_i$, $\omega_i$) per $Cr^{3+}$ ion and the parity combinations of vectors $D, P, Q_i, \Omega_{a,b}$ transforming under the action of the generators of space group *Pnma*. Note that the axial order parameters $\Omega_{a,b}$ were constructed accounting the antirotation of the octahedrons surrounding neighboring $Cr^{3+}$ ions numbered as 1,2 and 3,4.

Table B. Possible arrangements of the distortive order parameters in $RCrO_3$

| Arrangement of octahedron axes $\omega_i \parallel n_{oi}$ | Order parameters Axial vectors | Parity combinations |
|---|---|---|
| 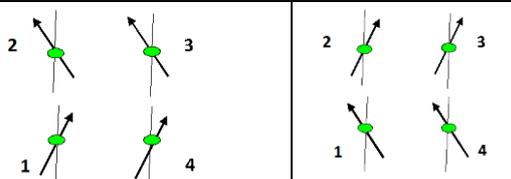 | $\Omega_a = \omega_1 - \omega_2 - \omega_3 + \omega_4$ $\bar{\Omega}_a = -\omega_1 + \omega_2 + \omega_3 - \omega_4 = -\Omega_a$ | $\bar{1}(+)2_x(+)2_y(-)2_z(-)$ |
| 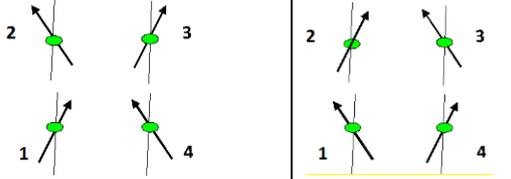 | $\Omega_b = \omega_1 - \omega_2 + \omega_3 - \omega_4$ $\bar{\Omega}_b = -\omega_1 + \omega_2 - \omega_3 + \omega_4 = -\Omega_b$ | $\bar{1}(+)2_x(-)2_y(+)2_z(-)$ |
| Arrangement of electric dipoles $d_i$ | Polar vectors | Parity combinations |
| 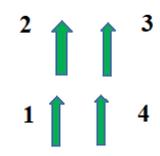 | $P = d_1 + d_2 + d_3 + d_4$ | $\bar{1}(+)2_x(+)2_y(+)2_z(+)$ |
| 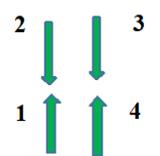 | $Q_2 = d_1 - d_2 - d_3 + d_4$ | $\bar{1}(+)2_x(+)2_y(-)2_z(-)$ |
| 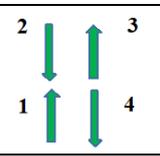 | $Q_3 = d_1 - d_2 + d_3 - d_4$ | $\bar{1}(+)2_x(-)2_y(+)2_z(-)$ |
| 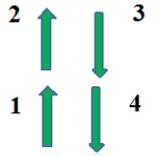 | $D = d_1 + d_2 - d_3 - d_4$ | $\bar{1}(+)2_x(-)2_y(-)2_z(+)$ |

**Appendix C**

*Table* C. Multiplications of IRs of *Pnma space* group

|    | Γ1 | Γ2 | Γ3 | Γ4 | Γ5 | Γ6 | Γ7 | Γ8 |
|----|----|----|----|----|----|----|----|----|
| Γ1 | Γ1 | Γ2 | Γ3 | Γ4 | Γ5 | Γ6 | Γ7 | Γ8 |
| Γ2 | Γ2 | Γ1 | Γ4 | Γ3 | Γ6 | Γ5 | Γ8 | Γ7 |
| Γ3 | Γ3 | Γ4 | Γ1 | Γ2 | Γ7 | Γ8 | Γ5 | Γ6 |
| Γ4 | Γ4 | Γ3 | Γ2 | Γ1 | Γ8 | Γ7 | Γ6 | Γ5 |
| Γ5 | Γ5 | Γ6 | Γ7 | Γ8 | Γ1 | Γ2 | Γ3 | Γ4 |
| Γ6 | Γ6 | Γ5 | Γ8 | Γ7 | Γ2 | Γ1 | Γ4 | Γ3 |
| Γ7 | Γ7 | Γ8 | Γ5 | Γ6 | Γ3 | Γ4 | Γ1 | Γ2 |
| Γ8 | Γ8 | Γ7 | Γ6 | Γ5 | Γ4 | Γ3 | Γ2 | Γ1 |